\documentclass[twocolumn,twoside,slac_two]{revtex4}
\usepackage{graphicx}
\usepackage{fancyhdr}
\begin{document}
\title{Review of Neutrino Oscillation Experiments}

\author{M.D.~Messier}
\affiliation{Department of Physics, Indiana University, Bloomington
  IN, 47405, USA}

\begin{abstract}
Several experiments have sought evidence for neutrino mass and mixing
via the phenomenon of neutrino flavor oscillations. In a three
neutrino model, these oscillations are described by three angles, two
mass splittings, and one CP violating phase. Experiments using
neutrinos from the Sun, the atmosphere, nuclear reactors, and particle
accelerators have gathered considerable information on these angles
and splittings. Two of the three angles are known to be large:
$\theta_{12} \simeq 33^\circ$, $\theta_{23} \simeq 45^\circ$, and an
upper limit is known on the third, $\theta_{13}<10^\circ$. Likewise,
the mass splittings are known to fall in the range $\Delta m^2_{12}
\simeq 8 \times 10^{-5}$ and $|\Delta m^2_{23}| \simeq 2.4 \times
10^{-3}$~eV$^2$. Several questions remain: the sign of the 2--3 mass
splitting, the size of the unknown angle $\theta_{13}$, and the size
of the CP violating phase are yet to be measured. Also, a report of
short-baseline $\bar{\nu}_e \rightarrow \bar{\nu}_mu$ oscillations has
yet to be confirmed. These open questions are the target of an
experimental neutrino oscillation program currently underway. This
report will attempt to summarize the current state of neutrino
oscillation measurements and the future program in as succinct a
manner as possible.

\end{abstract}

\maketitle

\thispagestyle{fancy}

\section{Introduction}
There is now in hand considerable evidence for neutrino flavor
oscillations, and hence neutrino mass and mixing. Neutrino
oscillations are determined by 6 parameters: two mass splittings,
$\Delta m^2_{12}$ and $\Delta m^2_{23}$, and 3 angles $\theta_{12}$,
$\theta_{23}$, $\theta_{13}$, and one CP violating phase $\delta$:
\begin{eqnarray}
\left[
\begin{array}{l}
\nu_e \\
\nu_\mu \\
\nu_\tau
\end{array}
\right]
&=&
\left[
\begin{array}{ccc}
1 &  & \\
 &  {\rm c}_{23} & {\rm s}_{23} \\
 & -{\rm s}_{23} & {\rm c}_{23}
\end{array}
\right]
\left[
\begin{array}{ccc}
{\rm c}_{13} &  & {\rm s}_{13}e^{-i \delta} \\
 & 1 &  \\
-{\rm s}_{13}e^{i \delta} &  & {\rm c}_{13}
\end{array}
\right]
\times \nonumber \\
&&
\hspace{25mm}
\left[
\begin{array}{ccc}
 \rm{c}_{12} & \rm{s}_{12} & \\
-\rm{s}_{12} & \rm{c}_{12} & \\
             &             & 1
\end{array}
\right]
\left[
\begin{array}{l}
\nu_1 \\
\nu_2 \\
\nu_3
\end{array}
\right]
\label{eq1}
\end{eqnarray}
Knowledge of the first and last of these matrices is derived from
measurements of solar neutrinos, reactor neutrinos, neutrinos from the
atmosphere, and neutrinos produced at accelerators. Currently, there is
no measurement which shows that the middle matrix is different from
unity and this matrix is the focus of a future program of
measurements. In this report, I will review the experimental
measurements of the parameters controlling neutrino oscillations.

\section{Current experimental status}
\subsection{$\theta_{12}$ and $\Delta m_{12}$}
Knowledge of the oscillation parameters $\theta_{12}$ and $\Delta
m^2_{12}$ come from observations of $\nu_e \rightarrow \nu_\mu +
\nu_\tau$ oscillations using neutrinos from the Sun and $\bar{\nu}_e
\rightarrow \bar{\nu}_\mu + \bar{\nu}_\tau$ using neutrinos from
nuclear reactors.

\begin{figure}[h]
\begin{center}
\includegraphics[width=80mm]{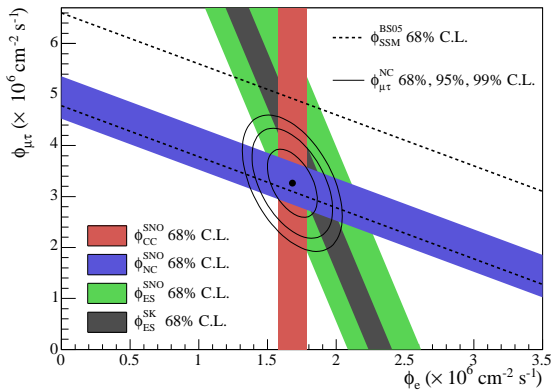}
\caption{Neutrino fluxes measured by the SNO and SK experiments. The
  exclusively CC and NC channels observed by SNO allow for extraction
  of the $\nu_e$ and non-$\nu_e$ components of the electron neutrino
  flux. These results are consistent with the measurements made by SK
  using CC and NC elastic scattering. The total neutrino flux is
  consistent with predictions from solar models. Reprinted
  from~\cite{Aharmim:2005gt}.
}
\end{center}
\label{solar-emutau-fluxes}
\end{figure}

The Sun produces an enormous flux of electron neutrinos ranging in
energy from a few keV up to several MeV in energy. These have been
detected on Earth by radio-chemical experiments including
Homestake~\cite{Cleveland:1998nv}, GALLEX~\cite{Cribier:1999am},
GNO~\cite{Altmann:2000ft,Altmann:2005ix}, and
SAGE~\cite{Abdurashitov:2002nt,Gavrin:2005ks} (see also the summary in
\cite{Cattadori:2005gq}) and by the real-time water Cherenkov
experiments Kamiokande, Super--Kamiokande
(SK)~\cite{Fukuda:1998fd,Fukuda:1998rq,Fukuda:1998ua,Fukuda:2001nj,
Fukuda:2001nk,Fukuda:2002pe,Smy:2003jf,Hosaka:2005um}, and the Sudbury
Neutrino Observatory
(SNO)~\cite{Ahmad:2001an,Ahmad:2002jz,Ahmad:2002ka,Aharmim:2005gt}.
Results of these experiments are summarized in Table~\ref{snu-table}.
\begin{table}[h]
\begin{center}
\caption{Summary of solar neutrino results. Rates are quoted in units
  of SNU's, fluxes in units of $10^{6}\nu/{\rm cm}^2/{\rm s}^2$.}
\begin{tabular}{l|lll}
\hline
Energy & Measurement & Expected \\
\hline
$>$0.233~MeV & $R=67.4^{+2.6}_{-2.3}$ & $127^{+12}_{-10}$\\
\multicolumn{2}{l}{~~GALLEX+GNO+SAGE} \\
\hline
$>$0.813~MeV  & $R=3.23 \pm 0.68$ & $8.2 \pm 1.8$ \\
~~Homestake \\
\hline
5-20~MeV & $\phi_{\rm ES}=2.35 \pm 0.02 \pm 0.08 $ & $5.79 \pm 1.33$ \\
~~SK     & $A_{\rm DN}^{\rm ES} = -0.021 \pm 0.020^{+0.013}_{-0.012}$ & 0\\
\cline{2-3}
~~SNO & $\phi_{\rm ES}=2.35 \pm 0.22 \pm 0.15 $ & $5.79 \pm 1.33$ \\
      & $\phi_{\rm tot}=4.94 \pm 0.21 ^{+0.38}_{-0.34} $ & $5.79 \pm 1.33$ \\
      & $A_{\rm DN}^{\rm ES} =  0.146 \pm 0.198 \pm 0.033 $ & 0 \\
      & $A_{\rm DN}^{\rm CC} = -0.056 \pm 0.074 \pm 0.053 $ & 0 \\
      & $A_{\rm DN}^{\rm NC} =  0.042 \pm 0.086 \pm 0.072 $ & 0 \\
\hline
\end{tabular}
\label{snu-table}
\end{center}
\end{table}
Each of these experiments observes a deficit of $\nu_e$'s relative to
expectations based on solar models
(eg.~\cite{Bahcall:2000nu,Bahcall:2004fg,Bahcall:2004pz,Turck-Chieze:2004mg}).
Confirmation that these deficits are due to a flavor-changing process
(ie. oscillations) by the SNO experiment. SNO uses 1~kt of D$_2$O
allowing separate measurements elastic ($\nu_x + e^- \rightarrow \nu_x
+ e^-$), charged-current ($\nu_e + d \rightarrow p + p + e^-$),
neutral-current ($\nu_x + d \rightarrow p + n + \nu_x$) scattering
rates. From these measurements, SNO has been able to confirm that the
total neutrino flux, $\phi_e + \phi_\mu + \phi_\tau$, from the Sun was
consistent with solar models and that the deficit of $\nu_e$'s was
compensated by a non-zero flux of $\nu_\mu$ $\nu_\tau$
(Figure~\ref{solar-emutau-fluxes}).

Interpretations of the deficits in terms of neutrino oscillations
historically fell into four categories in the mass-splitting-mixing
parameter space: vacuum oscillations (``VAC'') $\Delta
m^2_{12}\simeq10^{-10}~{\rm eV}^2$, ``LOW'' $\Delta
m^2_{12}\simeq10^{-7}~{\rm eV}^2$, small mixing angle (``SMA'') $\Delta
m^2_{12}\simeq10^{-5}~{\rm eV}^2$, $\tan^2\theta_{12}\simeq10^{-3}$,
and large mixing angle (``LMA'') $\Delta m^2_{12}\simeq 10^{-5-4}~{\rm
eV}^2$ $\tan^2\theta_{12}\simeq0.4$. Each region has its own expected
signatures: vacuum oscillations should produce an annual variation as
the Earth-Sun distance varies throughout the year, the small-mixing
solution should produce a significant spectral distortion in the
energy region below 5~MeV; in many cases there is expected to be a
significant matter effect from the Earth resulting in a day-night flux
asymmetry. A preference for the LMA solution began to emerge from the
Super--Kamiokande data which saw no significant spectra distortion of
the recoil electron energy spectrum and no significant day-night
asymmetry -- a trend which was strengthened by the SNO
measurements. Note that as the LMA solution produces a large matter
effect on the oscillations in the Sun, the sign of the 1--2 mass
splitting is determined to be positive by the solar neutrino data.

\begin{figure}[h]
\bigskip
\begin{center}
\includegraphics[width=80mm]{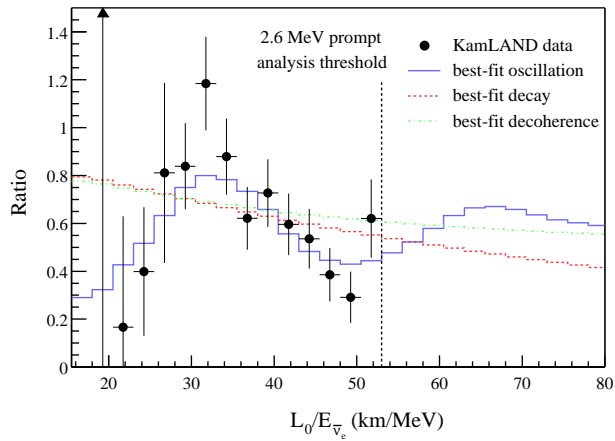}
\caption{The KamLAND event rate relative to non-oscillated
  expectations as a function of reconstructed $L/E$. The solid curve
  is for LMA oscillation parameters. Dashed curves show
  non-oscillation models and are shown to give some indication as to
  the significance of the dip near 50~km/MeV. Reprinted
  from~\cite{Araki:2004mb}
}
\label{kamland-loe}
\end{center}
\end{figure}

The validity of the LMA interpretation of the solar neutrino fluxes
was demonstrated conclusively by the KamLAND
experiment~~\cite{Eguchi:2002dm,Araki:2004mb}. KamLAND uses 1~kt of
liquid scintillator located in the former Kamiokande cavern to observe
$\bar{\nu}_e$'s from over 50 nuclear reactors located throughout Japan
and Korea via inverse beta decay. The majority of the neutrino flux
(79\%) comes from 26 reactors located at distances ranging from
138-214~km resulting in an average distance of 180~km. The long
baseline coupled with the low neutrino energy (~10--50 MeV) allows
KamLAND to test the solar LMA solution in a terrestrial
experiment. KamLAND observes a deficit of neutrinos who's distribution
in $L/E$ is consistent with LMA oscillations
(Figure~\ref{kamland-loe}).
The parameters favored by the solar neutrino and KamLAND data are not
only consistent with each other, but complement each other as the
solar neutrino observations are mostly sensitive to the mixing
parameter and the KamLAND measurements are most sensitive to the
mass-splitting. 
\begin{figure}[h]
\begin{center}
\includegraphics[width=80mm]{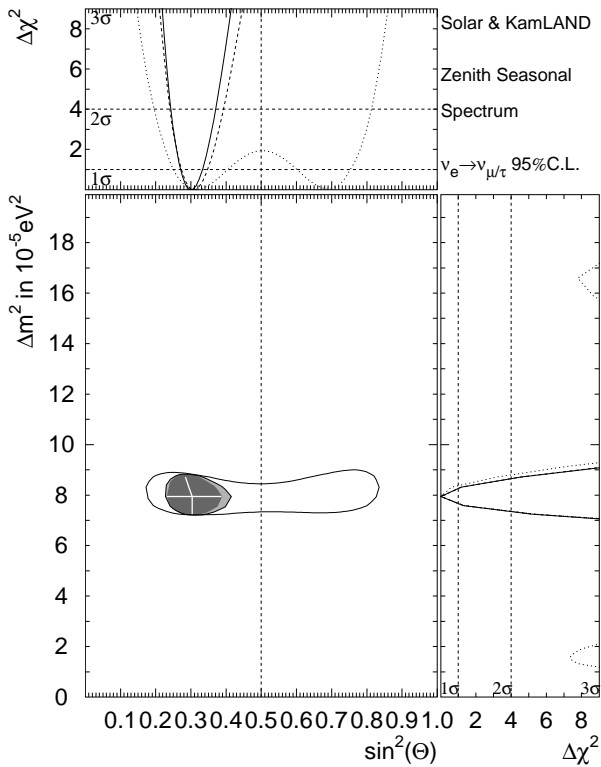}
\caption{ The allowed values of $\sin^2\theta_{12}$ and $\Delta
m^2_{12}$ at 95\% C.L.  The solid contour is for the KamLAND data
alone. The light gray region adds solar neutrino data from SNO and SK
and the dark gray region adds data from radio-chemical
experiments. Projections of the $\Delta \chi^2$ surfaces onto the
horizontal and vertical axes are shown at the top and side. Reprinted
from~\cite{Hosaka:2005um}.  }
\label{solar-kamland}
\end{center}
\end{figure}
Figure~\ref{solar-kamland} summarizes the regions of $\theta_{12}$ and
$\Delta m^2_{12}$ favored by the combined solar and KamLAND data.

\subsection{$\theta_{23}$ and $|\Delta m^2_{23}|$}
\begin{figure*}[th]
\begin{center}
\includegraphics[width=34mm]{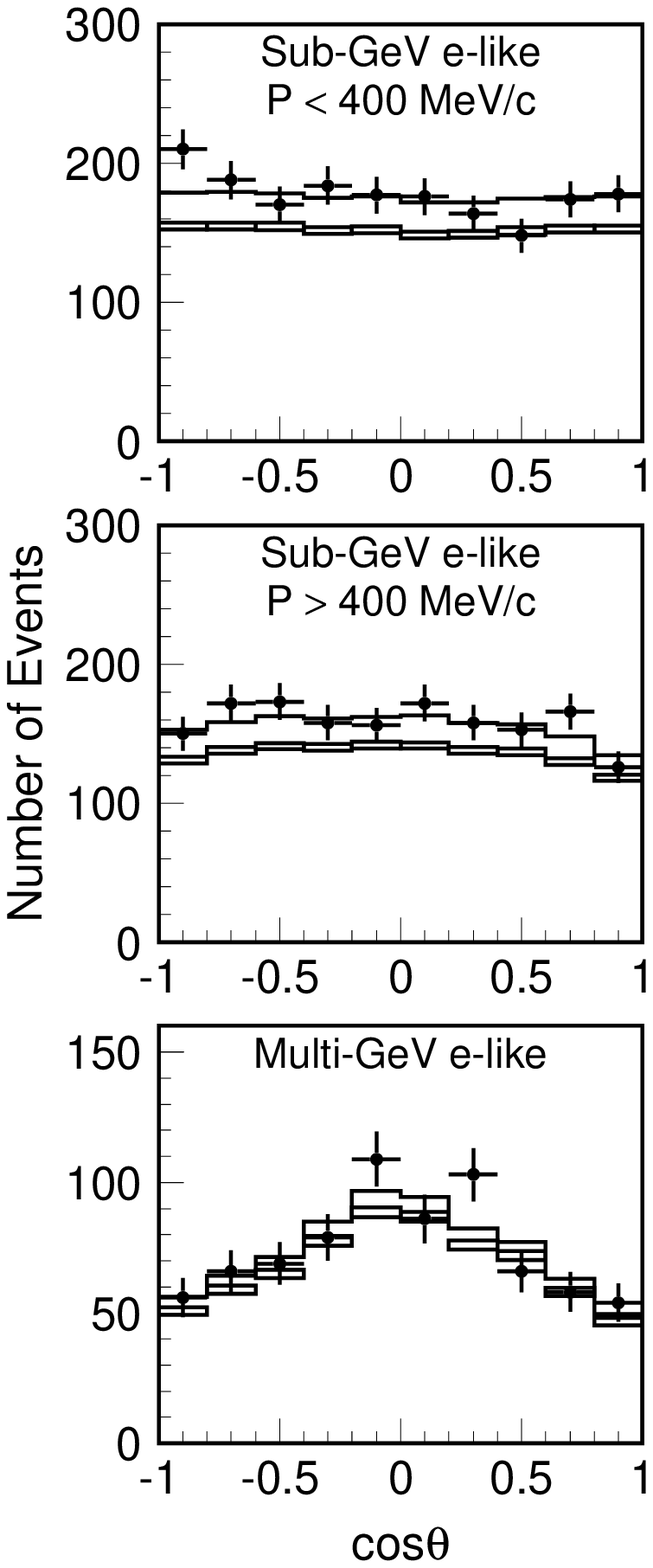}~
\includegraphics[width=63mm]{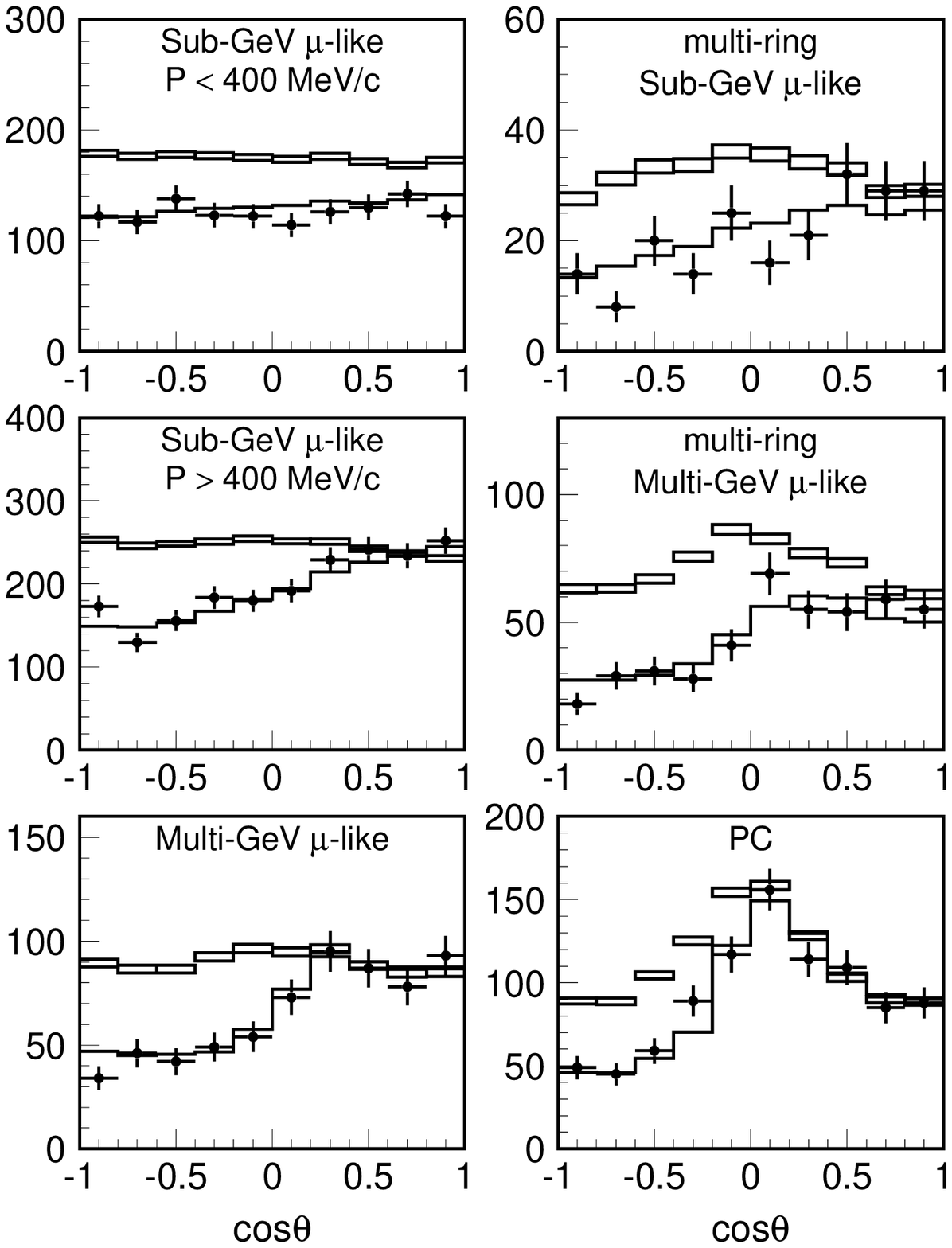}~
\includegraphics[width=64mm]{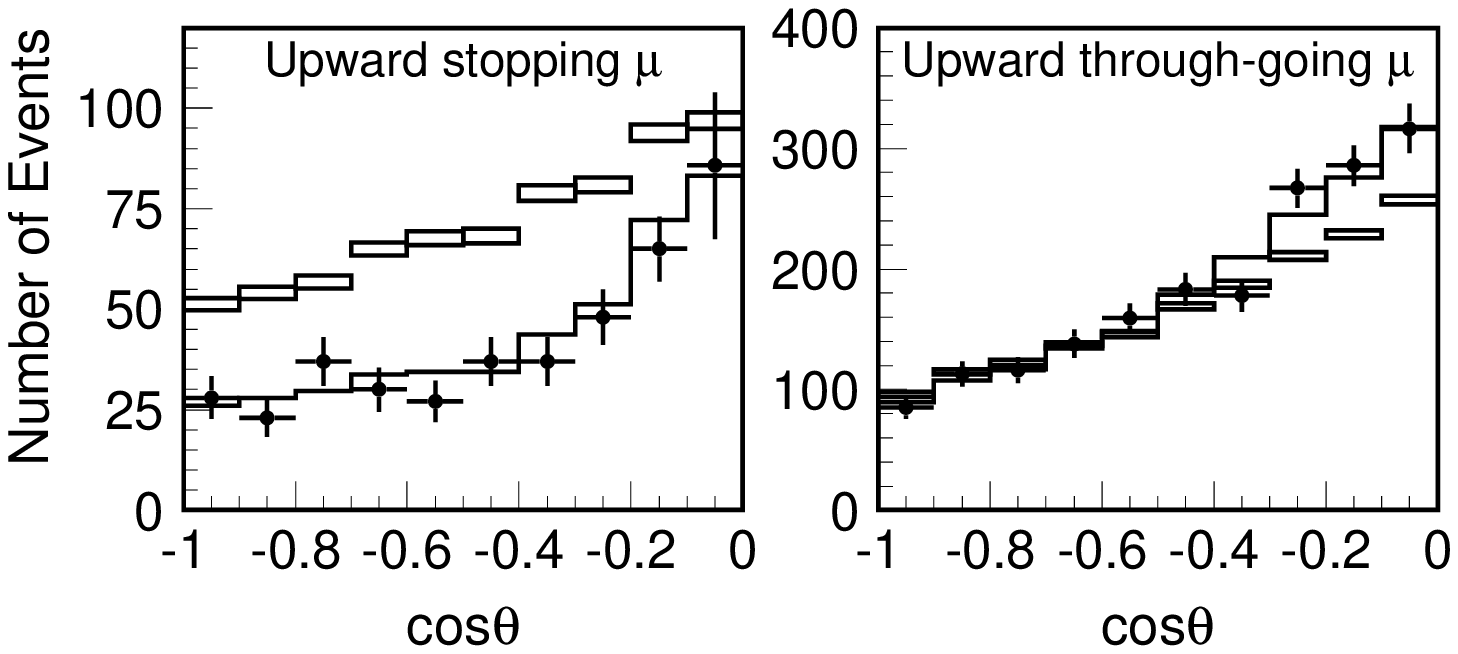}
\caption{Zenith rates of atmospheric neutrinos observed by SK. The
  left most panels show the electron neutrino rates as a function of
  energy; central panels show the contained and partially-contained
  muon neutrino event rates, and the right most panels show the upward
  stopping and upward through-going muon rates. In each case, the data
  is shown by points, the expectations without oscillations are shown
  by boxes, and the best-fit oscillated rates are shown by a single line.
}
\label{sk-atmnu-1}
\end{center}
\end{figure*}

Atmospheric neutrinos are produced in cascades initiated by cosmic-rays
collisions with nuclei in the Earth's atmosphere. The largest
production mechanism is $\pi^+ \rightarrow \mu^+ + \nu_\mu$, $\mu^+
\rightarrow e^+ + \nu_e + \bar{\nu}_\mu$ and charge-conjugates. While
absolute rates of atmospheric neutrino production have large $(\simeq
20\%)$ uncertainties, the relative rates of $\nu_e$ and $\nu_\mu$ can
be predicted with 5\% accuracy and the fluxes are expected to be
up/down symmetric with respect to the detector horizon. Several
experiments have observed atmospheric
neutrinos~\cite{Ambrosio:2004ig,Sanchez:2003rb,Gallagher:2005gx},
however, few experiments rival the high statistics of the SK
experiment. SK has collected contained $\nu_e$ and $\nu_\mu$ events
ranging in energy from 100~MeV through
20~GeV~\cite{Fukuda:1998tw,Fukuda:1998ub,Fukuda:1998mi} and
upward-going neutrino-induced muons ranging in energies from 20~GeV to
100~GeV~\cite{Fukuda:1998ah,Fukuda:1999pp}. This data set, which spans
roughly four orders of magnitude in neutrino energy, exhibits a
significant zenith-angle dependent deficit of $\nu_\mu$'s which is
well described by neutrino
oscillations~\cite{Ashie:2005ik}. Additionally, SK has isolated a
high-resolution data sample which shows hints of an oscillatory $L/E$
distribution~\cite{Ashie:2004mr}. Fits to this data yield results in
the range $1.5 \times 10^{-3} < |\Delta m^2_{23}| < 3.4 \times
10^{-3}~{\rm eV}^2$, and $\sin^2 2\theta_{23}>0.92$ (90\% CL).

\begin{figure}[h]
\begin{center}
\includegraphics[width=80mm]{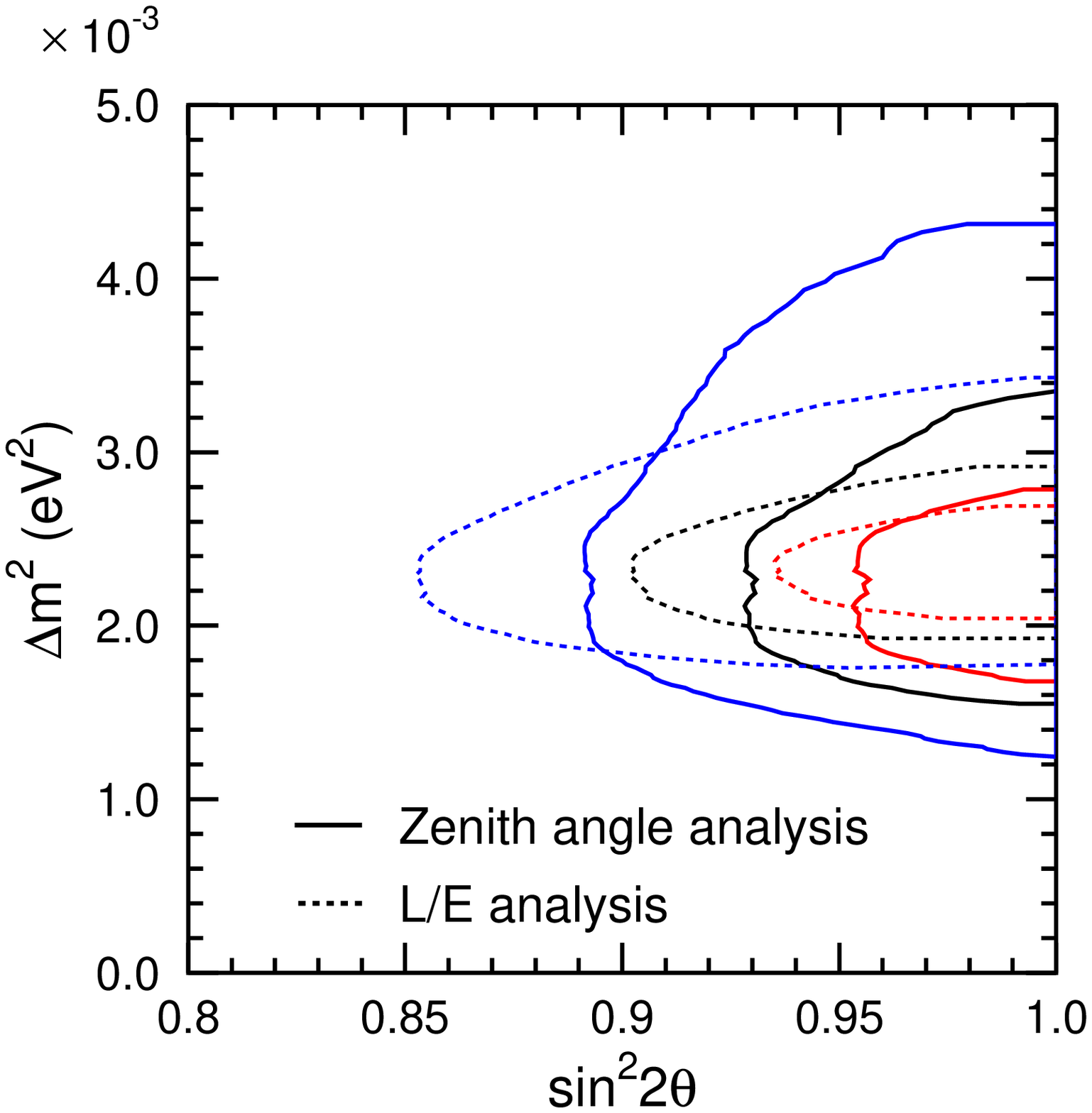}
\caption{Allowed parameter region from the SK atmospheric neutrino
results. Results are shown separately for the zenith-angle analysis
and the high-resolution $L/E$ analysis.
}
\end{center}
\end{figure}

\begin{figure}[ht]
\includegraphics[width=80mm]{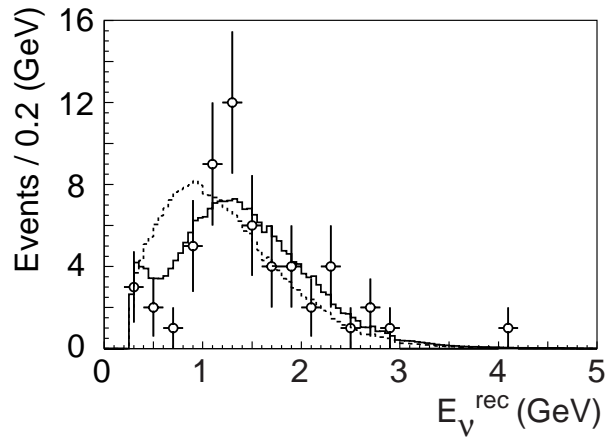}
\caption{The muon neutrino spectrum observed by the K2K experiment.}
\label{k2k-numu-spect}
\end{figure}

The atmospheric neutrino results obtained by Super--Kamiokande have
been confirmed by the K2K
experiment~\cite{Aliu:2004sq,Ludovici:2006gu}. K2K uses a ~98\% pure
beam of $\nu_\mu + \bar{\nu}_\mu$ of mean energy 1.3~GeV produced at
the KEK 12~GeV PS. The beam is directed at the SK detector a distance
of 250~km from the source. The experiment ran between 2001 and 2005
collecting a total of $1.049 \times 10^{20}$~POT.  The experiment has
recorded 112 events with an expectation of 159 before oscillations --
a $4.2~\sigma$ deficit. From fits to the energy spectrum of the 58
events which have a single muon (see Figure~\ref{k2k-numu-spect}), K2K
extracts a measurement of the oscillation parameters $\sin^2
2\theta_{23}>0.56$ and $|\Delta m^2_{23}|$ in the range from $1.88$ --
$3.48 \times 10^{-3}~{\rm eV}^2$ (90\% CL), in good agreement with the
SK atmospheric neutrino results.

Recently, the MINOS experiment has completed its first year of running
with the NuMI neutrino beam from Fermilab. During this run, MINOS
accumulated over $10^{20}$ protons on target and currently has enough
data to improve on SK's measurement of $\Delta m^2_{23}$. Details of
this new measurement are contained in these proceedings~\cite{tagg}.

\subsection{$\theta_{13}$}
\begin{figure}[h]
\begin{center}
\includegraphics[width=80mm]{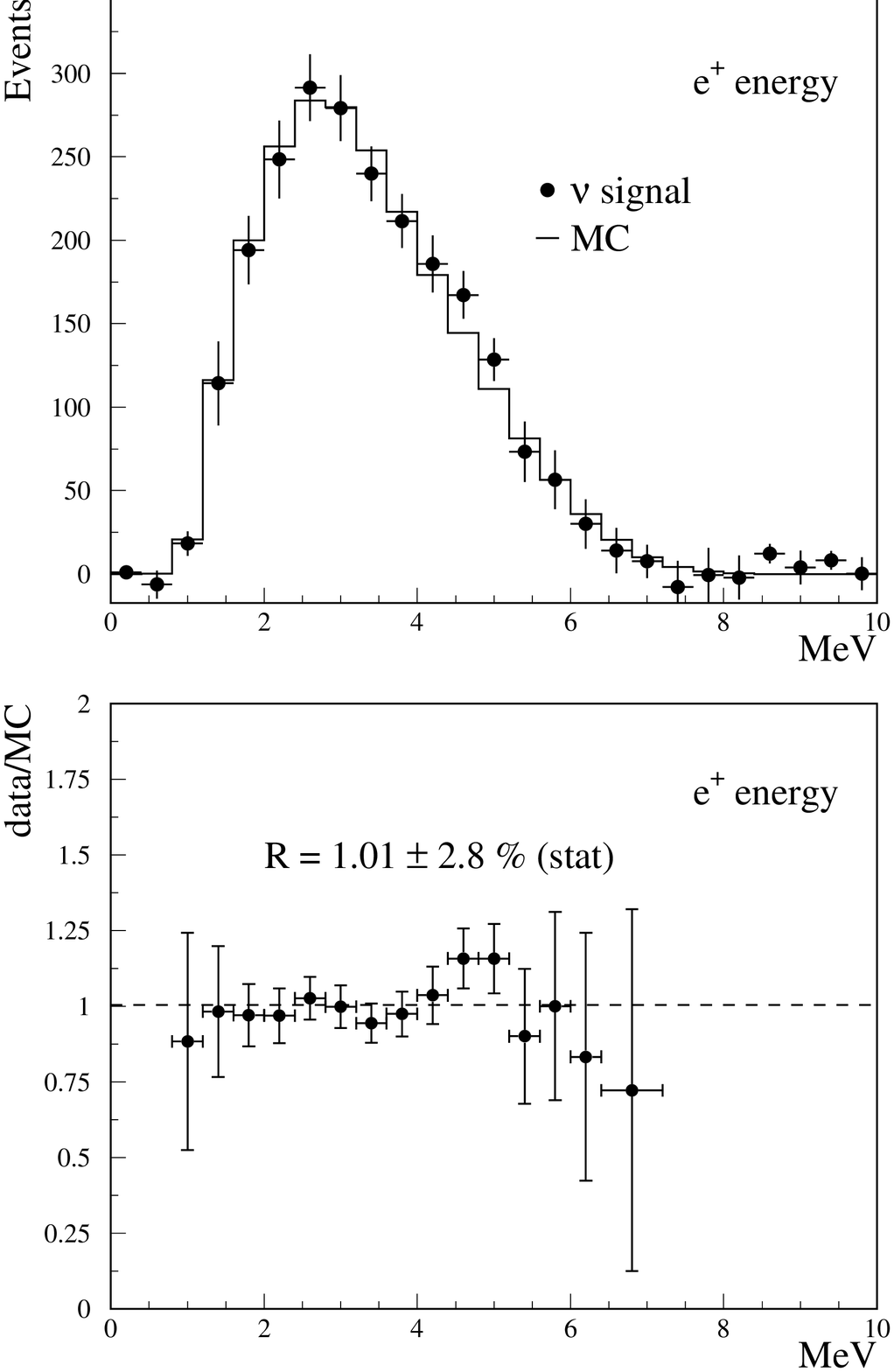}
\caption{The comparison of the expected positron spectrum and
  the observed spectrum in the CHOOZ experiment. Top is the rate
  distribution and bottom shows the ratio.}
\end{center}
\label{chooz-spect}
\end{figure}

Both solar and atmospheric oscillations show evidence for large
neutrino mixing. One might also expect, then, that the remaining
mixing angle, $\theta_{13}$ would also be large. However, to date no
observation of oscillations involving this angle have been made. The
most sensitive search has been made by the CHOOZ
experiment~\cite{Apollonio:2002gd} which looked for evidence of
$\bar{\nu}_e$ disappearance at the $\Delta m^2_{23}$ scale. The
comparison of the measured to the expected positron spectrum is shown
in Figure~\ref{chooz-spect}.
No evidence is seen for an oscillation and CHOOZ has set an upper
limit on $\sin^2 2\theta_{13}$ randing from 0.10 at the upper end of
the $\Delta m^2_{23}$ range indicated by atmospheric neutrinos to 0.15
at the lower end of that range. The CHOOZ results have been confirmed,
although with less sensitivity, by the K2K experiment which has looked
for $\nu_e$ appearance in their $\nu_\mu$
beam~\cite{Yamamoto:2006ty}. They find one event with an expected
background of 1.7 events setting a limit of roughly $\sin^2
2\theta_{13}<0.26$. Recently, SK has examined their multi-GeV electron
neutrino data for evidence of matter-enhanced oscillations in a search
for non-zero $\theta_{13}$~\cite{Hosaka:2006zd}. No evidence is found,
placing a limit on $\sin\theta_{13}<0.06$.

\subsection{LSND and miniBooNE}
In 1996 the LSDN collaboration reported evidence for appearance of
$\bar{\nu}_e$ in a $\bar{\nu}_\mu$ beam produced via muon decay in
flight and at rest~\cite{Athanassopoulos:1997pv,
Athanassopoulos:1997er,Athanassopoulos:1996jb}.  This result was not
confirmed KARMEN, a similar, though somewhat less sensitive
experiment~\cite{Armbruster:2002mp,Drexlin:2003fc}. The short baseline
of the LSND experiment, coupled with the relatively low neutrino
energies ($\simeq$10-50~MeV) suggests that these oscillations are
associated with a mass-splitting on the order of 1~eV$^2$. This
splitting is difficult to reconcile with the atmospheric and solar
neutrino oscillations which indicate a mass splitting more that two
orders of magnitude smaller. Attempts to explain the solar and
atmospheric neutrino oscillations and include the report from LSND
typically rely on extensions to the standard model including models
with a fourth, light, sterile, neutrino or CPT
violations. Confirmation of the LSND result would be a major
revolution in neutrino physics and is being pursued by the miniBooNE
experiment at Fermilab~\cite{Stancu:2006gv}.

\section{Future experiments: $\theta_{13}$,
 sign of $\Delta m^2_{23}$, and $\delta_{{\rm CP}}$
}

The future neutrino oscillation program seeks as its ultimate goal
evidence for CP violation in the lepton sector. As can be seen from
Eq.~\ref{eq1}, any CP violation enters into the neutrino mixing matrix
proportional to $\sin\theta_{13}$. Since there is currently only an
upper limit on this mixing parameter it is the focus of the next round
of neutrino oscillation measurements to be carried out at reactors and
accelerators.

\subsection{Future experiments at reactors}
There is current great interest in pushing the measurement technique
used by the CHOOZ experiment to gain roughly an order of magnitude
more sensitivity to $\sin^2 2\theta_{13}$. These include the
Double-CHOOZ~\cite{Ardellier:2004ui} experiment,
KASKA~\cite{Kuze:2005kz}, and Daya Bay~\cite{Cao:2005mh}
experiments. The main improvement sought by each of these experiments
is to make relative measurements between identical (or nearly
identical) detectors located at different distances from the reactor
core to cancel uncertainties in the absolute neutrino production
rates. The experiments expect to reach a sensitivity to $\sin^2
2\theta_{13}$ down to roughly 0.01. As these experiments measure
$\sin^2 2\theta_{13}$ via s disappearance channel, they are
insensitive to the affects of the CP violating phase $\delta$.

\subsection{Future experiments at accelerators}
Two experiments are going forward to search for electron neutrino
appearances in a muon neutrino beam. In Japan, a new neutrino beamline
is under construction at the 50~GeV PS at J-PARC which is directed at
the SK detector 295~km away for the T2K
experiment~\cite{Itow:2001ee}. In the first phase of the experiment is
expected to begin in 2009 with a beam intensity of 100~kW ramping up
to 0.9~MW by 2011. In its first run, T2K expects to have sensitivity
to $\sin^2 2\theta_{13}$ down to roughly 0.006 (90\% CL). Future
upgrades include an increase in the beam intensity to 4~MW and
construction of a new mega-ton scale water Cherenkov detector. With
these upgrades, it will be possible to begin to study of CP violation.

In the US, the NOvA~\cite{Ayres:2004js} experiment plans to construct
a new 25~kt scintillator tracking calorimeter at a distance 810~km
from the existing NuMI beam line. In its first run, NOvA plans to run
3 years in neutrino mode, and 3 year in anti-neutrino mode yielding a
sensitivity to $\sin^2 2\theta_{13}$ down to roughly 0.008
(2$\sigma$). Due to its long baseline, NOvA is sensitive to the sign
of $\Delta m^2_{23}$ and can begin to study the question of the mass
hierarchy in its first run. Later upgrades are imagined for NOvA,
including the possibility of a multi-kt liquid Argon detector located
at the second oscillation maximum and upgrades of the proton source
increasing the reach of the mass hierarchy measurement and opening
the possibility of searches for CP violation. Due to the large
difference in baselines (295~km vs. 810~km), the combination of the
data from T2K and NOvA greatly extend the search for CP violation
beyond what can be accomplished by one experiment working alone.

\bigskip 

\end{document}